\documentclass[multphys,vecphys]{svmult}

\usepackage{makeidx}         
\usepackage{graphicx}        
\usepackage{multicol}        
\usepackage[bottom]{footmisc}

\makeindex             

\begin{document}

\title*{Integral Field Spectroscopy of a peculiar Supernova Remnant MF16
in NGC6946}
\titlerunning{MF16}
\author{Abolmasov P.\inst{1,2}
Fabrika S.\inst{2}, Sholukhova O.\inst{2} \and Afanasiev V.\inst{2}}
\authorrunning{Abolmasov et al.}
\institute{Moscow State University 
\and
Special Astrophysical Observatory RAS}

\maketitle

\begin{abstract}
We present a study of a peculiar Supernova Remnant MF16, associated with the
Ultraluminous X-ray Source (ULX) NGC6946 ULX-1. Observations were taken with 
the MultiPupil Fiber Spectrograph (MPFS) with the 6-m telescope on January 2005.
The nebula is found to be highly asymmetric, one of the parts being much
denser and colder. 
Two-component structure of the emission lines and radial velocity
gradient in some of them argue for a non-spherical nebula, expanding
with a velocity of about $100km\,s^{-1}$.
Neither shock models nor the X-ray emission can adequately explain the 
actual emission line spectrum of MF16, so we suggest an additional 
ultraviolet source with a luminosity of about $10^{40} erg\,s^{-1}$.
We confirm coincidence of the ULX with the central star, and identify radio 
emission observed by VLA with the most dense part of the nebula.
\end{abstract}

\section*{Introduction}

The attention to MF16 was first drawn by Blair \& Fesen \cite{abol_BF_94},
who identified the object as a Supernova Remnant (SNR), according to the 
emission-line spectrum with bright collisionally-excited lines. 
It was long considered as a peculiar and unusually luminous SNR, because of its huge 
optical emission-line ($L_{H\alpha} = 1.9\times10^{39}erg\,s^{-1}$, according to \cite{abol_BF_94},
for the tangential size $20\times34pc$)
and X-ray ($L_X = 2.5\times10^{39}erg\,s^{-1}$ in the $0.5-8$keV range,
according to \cite{abol_RoCo}) luminosities.

However, it was shown by Roberts \& Colbert \cite{abol_RoCo} that
the spectral, spatial and timing properties of the X-ray source do 
not agree with the suggestion of a bright SNR, but rather suppose
a point source with a typical ``ULX-like'' X-ray spectrum: cool 
Multicolor Disk (MCD) and a Power Law (PL) component. So, apart from the 
physical nature of the object, MF16 should be considered a 
{\it ULX nebula}, one of a new class of objects, described 
by Pakull \& Mirioni, \cite{abol_Pakull}. For introduction into 
ULX-counterparts see \cite{abol_garching_Fabrika}.

\section*{Observations and Data Reduction}

We analyse spectra obtained with the SAO 6m telescope using Multipupil
Fiber Spectrograph (MPFS, \cite{abol_MPFSafan}), 
with a spectral resolution 
$\Delta\lambda \simeq 8 \AA$ in $4000-7000\AA$ spectral range. 
MPFS field consists of $16\times16$ $1'' \times 1''$ spaxels.
For the nebula size (in its brightest part) 
$1'' \times 1.''5$ it means rather poor spatial resolution,
yet one can resolve the structure of the object, considering offset pixels,
were the PSF wings of the closest parts of the nebula contribute. 

Data reduction system was written in IDL environment, using procedures,
written by V. Afanasiev, A. Moiseev and P. Abolmasov. We also added
atmospheric dispersion correction in order to culculate correctly the 
nebula barycenters.

\section*{Results}

More complete version of the results will be published in \cite{abol_MF16_main}.
Integral fluxes, velocities, FWHMs and barycenters for selected lines are 
listed in Table~\ref{tab:abol_lines}. The integral unreddened flux in
$H_\beta$ line is $F_{H\beta} = (2.00\pm0.06) 10^{-14}erg\,cm^{-2} s^{-1}$.
Assuming $H_\alpha / H_\beta = 3$ we find interstellar absorption 
$A_V = 1.^m24$. Unreddening was produced using Cardelli et al.~\cite{abol_CCM}
algorythm. In the Table~\ref{tab:abol_lines} we present the 
most important parameters of selected emission lines, including 
barycenter shifts versus {\it Chandra} source coordinates~\cite{abol_RoCo}. An interesting
result can be seen that the barycenter is shifted to the West for low-excitation lines
(Balmer lines, [SII], [OI]), and to the East for the most high-excitation line
HeII$\lambda$4686. That can be understood as a global physical conditions gradient:
the western part of MF16 is denser and colder than the eastern. 

\begin{table}
\caption{MF16 selected emission line properties.}
\begin{center}
\begin{tabular}{lcccccc}
$line$ & $F(\lambda)/F(H_\beta)$ & $F(\lambda)/F(H_\beta)$ & $V,km/s$
& $FWHM, \rm \AA$ & $\Delta\alpha_c, arcsec$ & $\Delta\delta_c, arcsec$ \\
 & & (unreddened) & & & \\
\hline
\noalign{\smallskip}
\hline
\noalign{\smallskip}
      $HeII\lambda$4686   &0.20$\pm$0.03 & 0.22$\pm$0.03  &-5$\pm$18    &6.8$\pm$0.7&0.302$\pm$0.05&0.146$\pm$0.03\\
      $H_\beta$           &1.00$\pm$0.03 & 1.00$\pm$0.03  &-28$\pm$4    &7.0$\pm$0.2&-0.11$\pm$0.02&-0.10$\pm$0.02\\
      $[OIII]\lambda$5007 &7.42$\pm$0.15 & 6.94$\pm$0.14  &-17$\pm$2    &6.7$\pm$0.1&0.058$\pm$0.015&0.011$\pm$0.015\\
      $[OI]\lambda$6300   &1.42$\pm$0.08 & 0.93$\pm$0.05  &25$\pm$9     &10.1$\pm$0.4&-0.196$\pm$0.015&-0.237$\pm$0.015\\
      $H_\alpha$          &4.73$\pm$0.09 & 2.96$\pm$0.06  &-16$\pm$2    &7.8$\pm$0.1&-0.111$\pm$0.013&-0.08$\pm$0.013\\
      $[NII]\lambda$6583  &4.16$\pm$0.08 & 2.59$\pm$0.05  &-19$\pm$2    &7.6$\pm$0.1&-0.061$\pm$0.013&-0.064$\pm$0.013\\
      $[SII]\lambda$6717  &2.46$\pm$0.02 &1.496$\pm$0.015 &-37.5$\pm$1.1&7.7$\pm$0.1&-0.116$\pm$0.014&-0.081$\pm$0.014\\
\noalign{\smallskip}
\hline
\end{tabular}
\end{center}
\label{tab:abol_lines}
\end{table}

\begin{table}
\caption{MF16 two-component emission lines properties}
\begin{center}
\begin{tabular}{lcccccc}
\hline
\noalign{\smallskip}
\hline
\noalign{\smallskip}

$\lambda, \rm \AA$ & $F_1(\lambda)/F(H_\beta)$ & $V_1,km/s$ & $FWHM_1$,$\rm \AA$ & $F_2(\lambda)/F(H_\beta)$ & $V_2,km/s$ & $FWHM_2$,$\rm \AA$\\

$HeII\lambda$4686 &0.107$\pm$0.007& -142$\pm$20 &11.6$\pm$0.6&0.093$\pm$0.004&
10.$\pm$5.&4.26$\pm$0.15\\
      
$H_\beta$ &0.424$\pm$0.037&-122$\pm$17&10.24$\pm$0.50&0.58$\pm$0.02&
-2$\pm$5&5.67$\pm$0.16\\

$[OIII]\lambda$4959&0.875$\pm$0.014&-122$\pm$3&8.5$\pm$0.1&1.486$\pm$0.012&
8$\pm$1&5.78$\pm$0.12\\
      
$[OIII]\lambda$5007&2.696$\pm$0.015&-115$\pm$1&8.77$\pm$0.04&4.455$\pm$0.012&
11$\pm$2&5.63$\pm$0.05\\
\noalign{\smallskip}
\hline
\end{tabular}
\end{center}
\label{tab:abol_twocomp}
\end{table}

For some of the line profiles, that seem to have distinct two-component structure,
we also produce two-gaussian fit (see Table~\ref{tab:abol_twocomp}).
The velocity shift between the line components is about $120km\,s^{-1}$ 
for all the lines resolved into two components. 
This value is consistent with the velocity gradient 
seen for [OIII]$\lambda$5007 line, exceeding $50km\,s^{-1}$ \cite{abol_MF16_main}
and directed along the major axis of the nebula.
Recent observations with higher spectral resolution \cite{abol_scart}
confirm the result, that the nebula expands with either $\sim120$ 
(if we suggest just one shock wave with precursor), or $\sim60 km\,s^{-1}$
(if we see the both receeding and approaching parts of the nebula) velocity.
In all the profiles the blue components are significantly broader,
corresponding to velocity dispersion $300-400km\,s^{-1}$.

Having the spectral line fluxes and kinematics, one can judge 
about the nature of the nebula. It is well-known 
\cite{abol_BFS,abol_RoCo} that the optical emission-line luminosity
of MF16 exceeds very much that of a SNR with comparable size, so one should 
suggest either a more powerful shock or an additional photoionizing source.
If we consider a pure shock-wave excitation,
the H$_\beta$ flux density will be connected to the shock parameters as 
was stated by Dopita \& Sutherland \cite{abol_DoSutI}:

$$
 \begin{array}{l}
F_{H\beta} = F_{H\beta,shock}+F_{H\beta,precursor} = 
  7.44 \times 10^{-6} \left(  \frac{V_s}{100 km\, s^{-1}} \right)^{2.41} 
\times \left( \frac{n_2}{cm^{-3}}\right)
+ \\
\qquad{} 9.86 \times 10^{-6} \left(  \frac{V_s}{100 km \,s^{-1}} \right)^{2.28}
\times \left( \frac{n_1}{cm^{-3}}\right) \, erg\, cm^{-2} s^{-1}
 \end{array}
$$

Here $n_1$ and $n_2$ are pre- and postshock number densisties, correspondingly.
Basing on this formula and assuming the nebula a sphere with 13pc radius,
and $n_2 \simeq 500 cm^{-3}$ \cite{abol_BFS}
one can see the potential shock velocity varies
from $150km\,s^{-1}$ (for $n_1 = 100cm^{-3}$, adiabatic shock) 
up to $350km\,s^{-1}$ (for $n_1 \to 0cm^{-3}$, fully radiative shock).

\section*{High-excitation lines}

[OIII] and HeII lines in MF16 appear to be unusually bright for a 
shock-ionized nebula. Observed ratio of
[OIII]$\lambda$5007 / H$_\beta \sim 7$ requires a shock 
velocity about $400km\,s^{-1}$ \cite{abol_evans}. 
As for HeII / H$_\beta \sim 0.2$ 
value, it cannot be explained by existing shock+precursor models
\cite{abol_evans,abol_DoSutI}

HeII ionization suggests a quite specific photoionizing source -- 
extreme ultraviolet (EUV) with $\lambda < 228 \AA$. An estimate of the EUV
luminosity can be made using ionizing quanta number estimate 
\cite{abol_osterbrock}:

$$
L_{\lambda < 228\AA} \ge  \frac{E_{UV}}{E_{\lambda4686}}
\frac{\alpha_B}{\alpha^{eff}_{HeII\,\lambda4686}}
\times L_{HeII\lambda4686} \simeq 10^{39} erg\,s^{-1}
$$

\section*{Photoionization Modelling}

We have computed a grid of CLOUDY96.01~\cite{abol_cloudy} photoionization models
in order to fit MF16 spectrum avoiding shock waves. We have fixed X-ray
spectrum known from Chanrda observations \cite{abol_RoCo} assuming
all the plasma is located at 10pc from the central point source,
and introduced a
blackbody source with the temperature changing from $10^3$ to $10^6$K and
integral flux densities from 0.01 to 100 $erg\,cm^{-2}\,s^{-1}$.

\begin{figure}
\centering
\includegraphics[width=12.5cm]{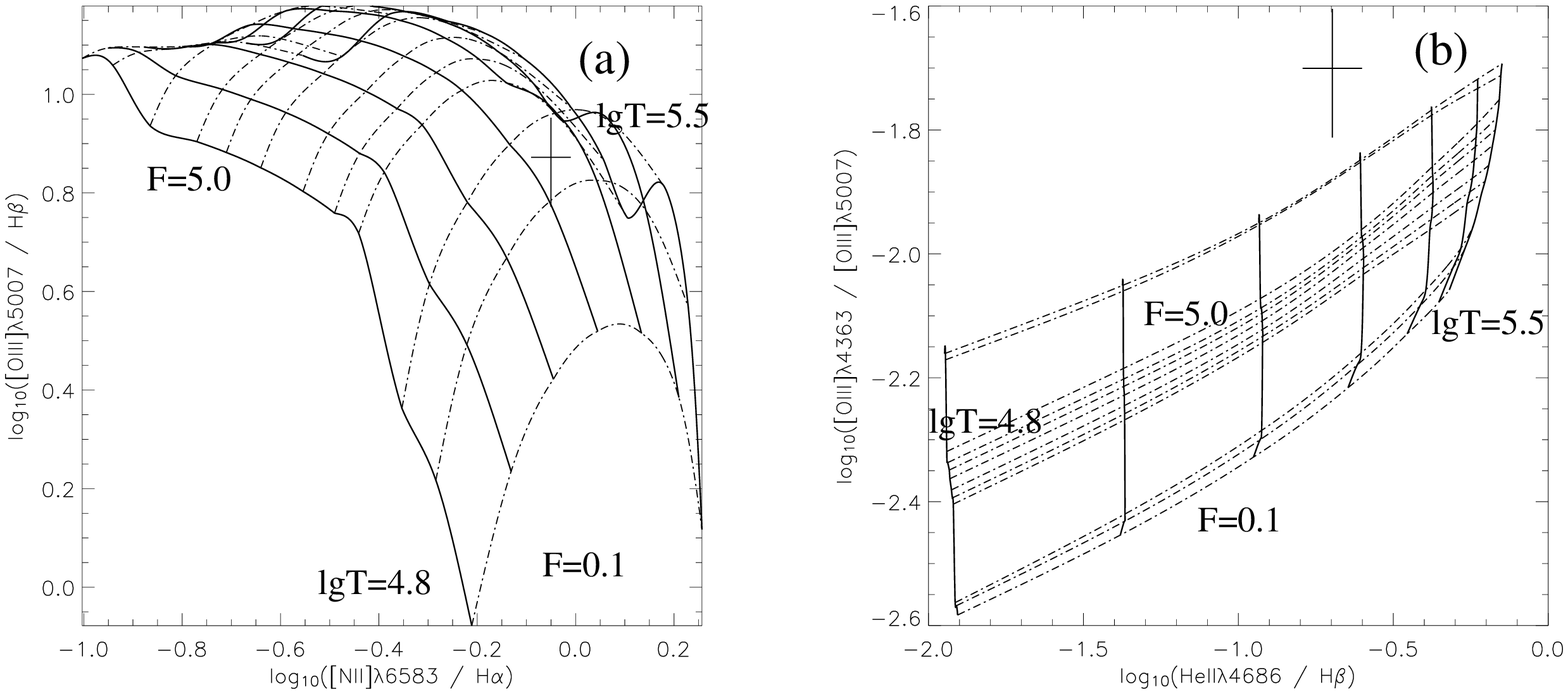}
\caption{$[OIII]\lambda$5007$ / H_\beta$  versus $[NII]\lambda$6583$ / H_\alpha$ (a)
and $HeII\lambda$4686 / $H_\beta$ versus $[OIII]\lambda$4363 / $[OIII]\lambda$5007 (b)
ionization diagrams with the integral spectrum of MF16
shown by a cross of error bars.
Photoionization grid descibed in the text is also shown. Solid lines trace
the constant temperature levels, dot-dashed lines correspond to constant
black body flux densities. Only models with $\lg T(K) = 4.80-5.30$
and $F = 0.1-7.0 erg\,cm^{-2}\,s^{-1}$ are shown.
}
\label{fig:abol_diag2}
\end{figure}

The best fit parameters are $\lg T(K) = 5.15\pm 0.05, F = 0.6\pm 0.1 erg\,cm^{-2}\,s^{-1}$,
that suggests quite a luminous ultraviolet source: $L_{UV} = (7.5\pm0.5) \times 10^{39} erg\,s^{-1}$.
More completely the modelling results and methodics
will be presented in \cite{abol_scart}. 
The UV source is more than 100 times brighter then what can be predicted by 
extrapolating the best-fit model for X-ray data \cite{abol_RoCo}.

\section*{Conclusions}

MF16 has a nontrivial emission line spectrum, similar to those 
observed in NLRs of Seyfert galaxies.
The nebula is highly asymmetric, with a dense cold western part, 
possibly connected with the radio-bright region \cite{abol_vandyk}.
Kinematical properties suggest an expanding non-spherical shell.
The sources of ionization acting in different parts of the nebula 
are probably different, strong radiative shock at the western end and in 
the outer regions, and hard UV/EUV radiation source in the 
inner/eastern regions.

Basing on the HeII4686 luminosity we should suggest a $\sim10^{39} erg\,s^{-1}$ EUV 
($\lambda \le 228\AA$) source responsible for the second ionization of He.
CLOUDY simulations suggest a $L_{UV}\sim10^{40} erg\,s^{-1}$ FUV 
luminosity needed to
produce the actual [OIII]$\lambda$5007 / $H_\beta$ ratio.

\bigskip

The research has been supported by the RFBR grants 04-02-16349 and 03-02-16341
and also by the joint RFBR/JSPS grant 05-02-19710.

\printindex
\end{document}